# Optimal-Control Suggestion for Congestion on Freeways using Data Assimilation of Distributed Fiber-Optic Sensing

**Yoshiyuki Yajima***
Senior Researcher
NEC Corporation
Kawasaki, Kanagawa, Japan, 211-8666
Email: yoshiyuki-yajima@nec.com, ORCiD: https://orcid.org/0000-0002-5476-9101

**Hemant Prasad**
Researcher
NEC Corporation
Kawasaki, Kanagawa, Japan, 211-8666

**Daisuke Ikefuji**
Professional
NEC Corporation
Kawasaki, Kanagawa, Japan, 211-8666

**Takemasa Suzuki**
Manager
NEC Corporation
Kawasaki, Kanagawa, Japan, 211-8666

**Shin Tominaga**
Director
NEC Corporation
Kawasaki, Kanagawa, Japan, 211-8666

**Hitoshi Sakurai**
Senior Professional
NEC Corporation
Kawasaki, Kanagawa, Japan, 211-8666

**Manabu Otani**
Deputy Manager
Central Nippon Expressway Company Limited
Nagoya, Aichi, Japan, 460-0003

*Corresponding Author

*Y. Yajima, H. Prasad, D. Ikefuji, T. Suzuki, S. Tominaga, H. Sakurai, and M. Otani*

**ABSTRACT**
This paper presents the optimal-control suggestion for congestion on freeways using data assimilation (DA) of distributed fiber-optic sensing (DFOS). To simultaneously maximize throughput and avoid/mitigate congestion, it is necessary to execute optimal control for the current traffic state as active transportation and demand management (ATDM) according to multi-objective optimization with real-time monitoring data. However, optimal control cannot be estimated due to intermittent observed data obtained from conventional sensors. To solve the issue, this paper proposes the ATDM optimal control estimation with DA of DFOS, which can monitor traffic flow in real time without dead zones. Our real-time DA method enables us to estimate the effectiveness of control scenarios by simulation. This paper also provides a method to uniquely determine the optimal-control solution among the Pareto solutions for multi-objective optimization. Throughput and mean speed across the entire road are considered as the objective functions. Variable speed limit (VSL) and inflow control are taken as ATDM examples. Validation results on a Japanese freeway show that (i) the optimal control scenario varies depending on the traffic state, especially congestion level; (ii) optimal control considering VSL alone improves throughput by 5–14% while the improvement rate for mean speed is 0–8%; (iii) throughput and mean speed are improved by 10–15% and 20–30%, respectively when VSL and inflow control are considered. This paper also implies the importance of balance management for the lane occupancy and proactive optimal control before congestion occurs.

*Y. Yajima, H. Prasad, D. Ikefuji, T. Suzuki, S. Tominaga, H. Sakurai, and M. Otani*

**INTRODUCTION**

Traffic congestion is a serious issue, particularly on freeways, because they are main infrastructure for transportation. When congestion level worsens, transportation efficiency decreases in addition to the increased accident risks and environmental burden from exhaust gas. To avoid or mitigate these negative impacts, it is necessary to execute optimal control within a minute or a few minutes for the latest traffic state estimated from the real time monitoring data. Specifically, throughput should be maximized to keep transportation efficiency high; on the other hand, traffic volume should be reduced to avoid or mitigate congestion. In such a case, optimal control is the solution of multi-objective optimization. Optimal control is carried out as active traffic management (ATM) and traffic demand management (TDM). ATM is a traffic management approach according to the current traffic state to prevent congestion and jam from occurring and growing. Representative implementations are variable speed limit (VSL), the part-time shoulder use, and the ramp metering (*1*). TDM is a management approach to keep traffic flow smooth and prevent congestion as it influences the behavior of drivers. For example, users' route choice is adjusted by the dynamic road pricing. In addition, TDM attempts to mitigate congestion by suggesting alternative transportation modes to users. ATM and TDM have recently been regarded as one concept: active transportation and demand management (ATDM).

Previous studies have reported various ATDM benefits. For example, VSL harmonizes the speed of traffic flow (*2*), which results in reduction of accident probability. VSL can also mitigate congestion because the number of vehicles reaching the bottleneck and the tail end of congestion decreases (*3, 4*). In addition, the critical density of the transition into congestion increases, which stabilizes traffic flow (*5, 6*). To control traffic demand, dynamic pricing is a common approach (e.g., *7*). When traffic demand is high, inflow to a congested road is indirectly restricted by increasing toll and drivers are encouraged to take another route. Inflow can also be controlled by optimizing ramp metering (e.g., *8*) and tollbooth operation (e.g., *9*) to avoid the disturbance of the mainline.

To reap these ATDM benefits, it is necessary to accurately predict the effectiveness of traffic control scenarios in simulations (*10*) and identify optimal control. However, the effectiveness is unpredictable because the current traffic state cannot be correctly reproduced in the traffic simulation. Its fundamental cause is that the conventional traffic sensors are unable to obtain the traffic state for a wide area in real time. Conventional traffic counters, such as surveillance cameras, loop inductors, and ultrasonic detectors, limit observable areas to 10–100 meters although they are usually installed at intervals of a few kilometers. Namely, most of road sections are dead zones. Probe-vehicle information also lacks real-time property due to the time lag of data uploading to roadside antennas, especially in congestion. Such spatiotemporal intermittent traffic data makes it challenging to model the current traffic flow in the simulation. Therefore, it is difficult to evaluate the effectiveness of traffic control scenarios. As a result, traffic operators do not have the solution of optimal control, such as when and how much to change the speed limit, which lane should VSL be applied to, and how much to reduce inflow through demand management. It is a fatal issue to manage and control traffic flow in real time so that traffic efficiency keeps as high as possible.

To overcome the issue, this paper proposes an estimation method of the optimal control scenario through data assimilation of distributed fiber-optic sensing (DFOS). DFOS is a sensing technology that detects and localizes vibration sources along fiber cables in real time using Rayleigh backscattering light of incident laser pulse from the apparatus to the optical fiber (*11*). DFOS obtains the trajectories of vehicles without dead zones in real time using already existing telecommunication fiber-cables along the road; thus, there is no need to install many additional sensors along the road. Since the light attenuation is suppressed in the optical fibers, DFOS can detect vibrations occurring at locations as far as 50–100 km away from the apparatus. Hence, DFOS has the advantage in terms of observable spatiotemporal coverage, real time property, installation and maintenance cost, and power saving compared with the conventional traffic sensors.

The application of DFOS for the transportation domain has been emerging. For example, it can be used for vehicle speed measurement and counting (*12–14*), vehicle type classification (*15*), pavement surface monitoring (*16, 17*), vehicle tracking (*18*), and vehicle-behavior recognition (*19*). Narisetty *et al*.

*Y. Yajima, H. Prasad, D. Ikefuji, T. Suzuki, S. Tominaga, H. Sakurai, and M. Otani*

(*20*) develop the data processing system for denoising of DFOS data to extract trajectories and visualize traffic flow. It derives spatiotemporally continuous mean speeds at the resolution of 0.5–1 km and 1 minute with the accuracy of 90% compared with the conventional traffic sensors. This system enhances the robustness of DFOS data-processing and enables us to monitor traffic flow in practical environments of freeways. It is already implemented on several freeways in Japan and utilized by road operators.

Taking advantage of the real-time property and wide coverage of mean speed obtained from DFOS, the authors have developed a data assimilation method for traffic simulation based on DFOS mean speed. Our data assimilation method models the current traffic flow and reproduces it in the simulation (*21–23*). The proposed method shows the accurate modeling results for the traffic state estimation with the accuracy of 80–97% (*21, 23*). Furthermore, it significantly reduces the prediction error of congestion by 70–80% compared with conventional methods (*22*). The accurate data assimilation motivates us to estimate optimal control for ATDM by simulating the effectiveness of traffic control scenarios.

Moreover, this paper presents an estimation method to uniquely determine the optimal control scenario in multi-objective optimization. In multi-objective optimization, Pareto solutions, which are the group of solutions that are trade-off each other, are generally obtained. This paper provides a way to derive the optimal solution among Pareto solutions by introducing the generalized distance with the consideration of users' requirements.

The objectives of this paper are as follows. (i) Using congestion data obtained on an actual freeway, this paper validates the optimal control estimation in multi-objective optimization based on our data assimilation method with DFOS. Two control scenarios are considered as ATDM examples based on the application of mean speeds by DFOS; one is only VSL and the other is the combination of VSL and inflow control as a generalized traffic demand control. The objective functions are throughput and mean speed, which reflect transportation efficiency and congestion level, respectively. (ii) This paper investigates which scenario is optimal control depending on the elapsed time since congestion occurs. This paper also considers two cases for objective orientation; one is a throughput-oriented and the other is a speed-oriented case.

**METHOD TO ESTIMATE OPTIMAL CONTROL**
**Figure 1** shows the overview of our proposed method with the ATDM Cycle advocated by the Federal Highway Administration, U.S. Department of Transportation (*1*). The spatiotemporally continuous traffic data by DFOS can promote the ATDM cycle more efficiently and accurately. In Step 1, data assimilation is carried out between mean speeds observed by DFOS and the traffic simulation. The brief process of data assimilation is as follows.

**Figure 2** shows the outline of our data assimilation method. In step 1.1, simulation data sets adopting multiple model-parameter sets are generated to emulate observed mean speeds by DFOS. Traffic volume observed by the traffic counter is used to run the simulations. In step 1.2, the posterior probability distributions (PPDs) of model parameters are estimated from the simulation data sets and observed mean speeds using the particle filter. PPDs are derived from the differences in the observed and simulated mean speed at each time and road segment. Based on the PPDs, the optimal values of model parameters reproducing the observed mean speeds in the simulation are identified. This study adopts the maximum a posteriori as the optimal value. In step 1.3, the latest microscopic traffic state, i.e., position and speed of each vehicle, is estimated from the latest mean speeds by DFOS and the fundamental diagram. It is used for the initial condition of the simulation from now on. In step 1.4, the simulation setting of the cell-automata model adopting the optimal model parameters and the microscopic traffic state is output. Details about the processes of the particle filter and the microscopic traffic state estimation are described in (*22*). This data assimilation enables the simulation to evaluate future traffic flow based on the current state.

In Step 2, traffic control scenarios are created and input into the simulation. In the case of VSL, for example, the speed limit of the road in the simulation is changed. When the control scenarios can be parametrized, several patterns of control scenarios are considered. For instance, this paper considers inflow control as demand management. In this case, the change rate of controlled inflow is a parameter. Among these control scenarios, the proposed method estimates the optimal control scenario regarding the condition where users prioritize in the latter step.

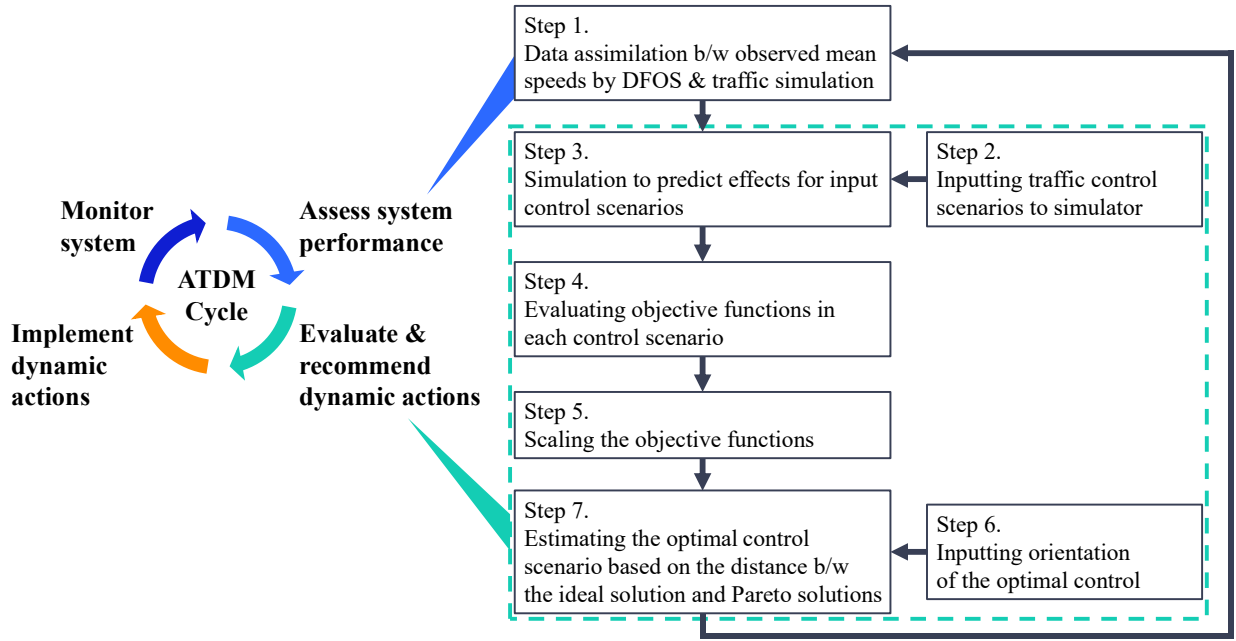

**Figure 1 Overview of the estimation process of the optimal control scenario. Step 1 and Step 2–7 correspond to the "assessment" and "evaluation & recommendation" steps, respectively, in the ATDM Cycle (*1*).**

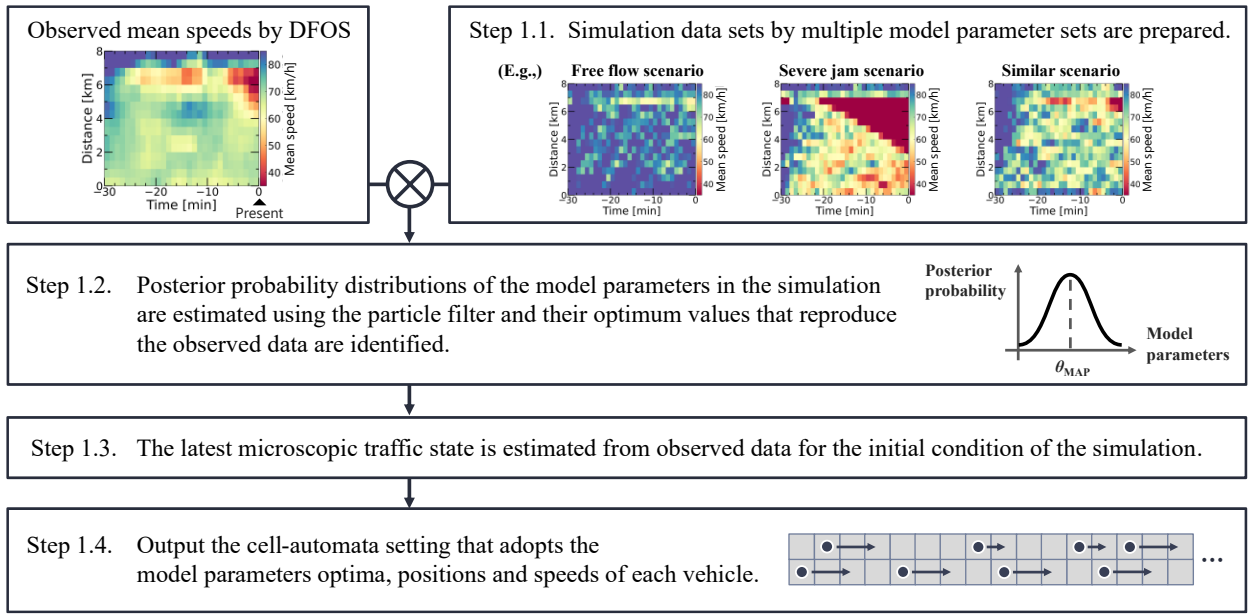

**Figure 2 Outline of data assimilation for the traffic simulation using mean speeds by DFOS.**

In Step 3, future traffic flow for each control scenario is simulated with the data assimilation results in Step 1. Specifically, the optimal values of the model parameters and the initial condition for the latest traffic state are adopted. Based on the simulation results, the objective functions in each control scenario are evaluated in Step 4. **Figure 3 (a)** shows simulation results on the objective-functions plane. Among these results, Pareto solutions are identified. They are candidates for the optimal control scenario.

This paper introduces processes to uniquely determine the optimal solution among Pareto solutions. The first process is Step 5: scaling the objective functions so that they have the same dimension and range. Objective functions in multi-objective optimization are usually heterogeneous; thus, it is necessary to match their ranges to equally evaluate them. For the scaling, this paper uses a standard value of the objective functions. When the objective functions are throughput and mean speed, the standard values can be the maximum flow $Q_{\max}$ derived from the speed–flow diagram and the speed at free flow $V_f$, respectively. Namely, the scaled throughput $\tilde{Q}_{TP}$ and scaled mean speed $\tilde{V}$ are defined as $\tilde{Q}_{TP} \equiv Q_{TP}/Q_{\max}$ and $\tilde{V} \equiv V/V_f$, respectively. This scaling process makes the objective functions dimension-less and have the same range $\tilde{Q}_{TP}, \tilde{V} \in [0, 1]$. Pareto solutions are re-distributed on the scaled objective-functions plane as shown in **Figure 3 (b)**.

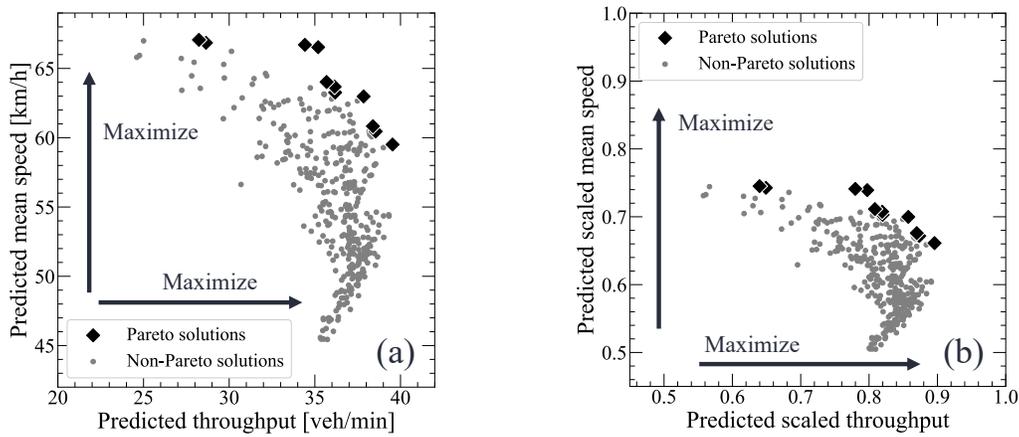

**Figure 3 (a) Distribution of Pareto (black) and non-Pareto (gray) solutions on the objective-functions plane. (b) The scaled objective functions of the panel (a).**

In Step 6, users' orientation for optimization is input and the optimal control scenario is estimated in Step 7. To do so, the ideal solution and the generalized distance from the ideal solution is introduced. The ideal solution is defined as the scaled objective functions when the system is the ideal but unfeasible state, for example, $(\tilde{Q}_{TP}, \tilde{V}) = (1, 1)$, i.e., maximum throughput at the free flow speed. The distance $d$ from the ideal solution is defined as the following equation using $L^p$-norm and the weight of an objective function,

$$d(w, p) = \sqrt[p]{\left(w|\tilde{Q}_{TP} - \tilde{Q}_I|\right)^p + \left[(1-w)|\tilde{V} - \tilde{V}_I|\right]^p}, \tag{1}$$

where $w$ is the weight of the scaled throughput $\tilde{Q}_{TP}$, $\tilde{Q}_I$ is the ideal solution of the scaled throughput, and $\tilde{V}_I$ is the ideal solution of the scaled mean speed $\tilde{V}$. In the case that there are three or more objective functions, **Equation 1** can be generalized as follows,

$$d(w, p) = \sqrt[p]{\sum_n (w_n|\tilde{f}_n - \tilde{f}_{n,I}|)^p}, \quad \left(\sum_n w_n = 1\right), \tag{2}$$

*Y. Yajima, H. Prasad, D. Ikefuji, T. Suzuki, S. Tominaga, H. Sakurai, and M. Otani*

where $\tilde{f}_n$ is the *n*-th scaled objective function, $\tilde{f}_{n,\text{I}}$ is the ideal solution of $\tilde{f}_n$, and $w_n$ is the weight of $\tilde{f}_n$. The optimal control scenario is estimated as the scenario that corresponds to the Pareto solution with the minimum $d(w,p)$. To define norm, $w$ and $p$ are input according to users' orientation for the optimization as described below.

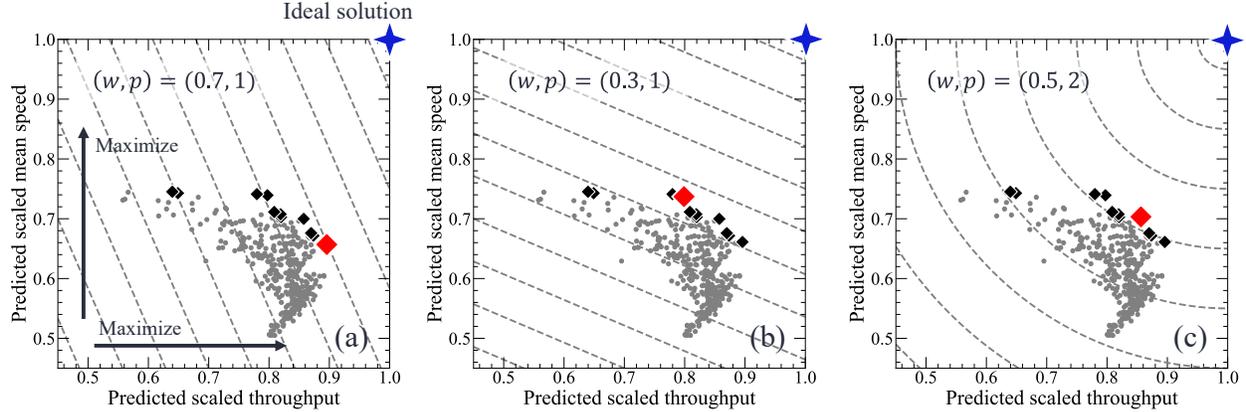

**Figure 4 Estimated optimal solution denoted as the red diamond for (a) throughput-oriented, (b) speed-oriented, and (c) balance-oriented scenarios. The ideal solution is denoted as the blue star. Dashed lines are contours of norm from the ideal solution at given $(w,p)$ denoted on the top left.**

The weight $w$ and the degree of norm $p$ in **Equation 1 and 2** control which objective function is prioritized and whether the optimal solution is oriented toward one objective function. In the case of the equal weight, $d$ at $p = 1, 2, \infty$ is known as the taxicab distance, Euclidean distance, and Chebyshev distance, respectively. **Figure 4 (a)–(c)** show contours of the distance at $(w,p) = (0.7, 1), (0.3, 1)$, and $(0.5, 2)$ and the optimal solution. When an uneven weight is adopted as **Figure 4 (a) and (b)**, the estimated optimal control scenario is oriented toward the weighted objective function. In contrast, the optimal solution in **Figure 4 (c)** is not oriented toward either objective function, i.e., a balance-oriented solution is selected. In this way, the optimal solution and the corresponding optimal control scenario are uniquely determined among the Pareto solutions considering the priority of objective functions. In practice, when there are no prioritized objective functions, the equal weight is set; otherwise, a higher weight is applied to a prioritized one. When users' orientation is balanced, two or a larger value is set to $p$; otherwise, it is set to be unity. For example, if users are oriented to optimize an objective function, but do not have the idea which objective function should be prioritized, they can set $(w,p) = (0.5, 1)$. This paper focuses on the throughput- and speed-oriented control scenarios where $(w,p)$ is $(0.7, 1)$ and $(0.3, 1)$, respectively.

**VALIDATION DATA & SIMULATION SETTING**
The optimal control estimation based on our data assimilation method is validated using traffic congestion data obtained on a part of Japanese freeway Route E85, Odawara-Atsugi Road, which is an access-controlled radial road from the Greater Tokyo Area and operated by Central Nippon Expressway Company (NEXCO Central) Limited. **Figure 5** shows spatiotemporal mean speeds on Route E85 obtained by DFOS. The fiber cable used in the measurement is buried at the depth of 60 cm beneath the shoulder for telecommunications between management offices of the freeway. The DFOS apparatus is installed in the electrical room located at the origin of the freeway. The validation section is 8 km out of the total 32 km long. At 7.1 km, there is a tollbooth as the bottleneck. It caused congestion with rapidly increased traffic volume. This paper refers to congestion when a road segment where the mean speed is below 40 km/h exists for 15 minutes according to the definition in NEXCO Central. Congestion finally reaches the maximum length and dissipates after 35 and 65 minutes have passed since congestion occurs, respectively.

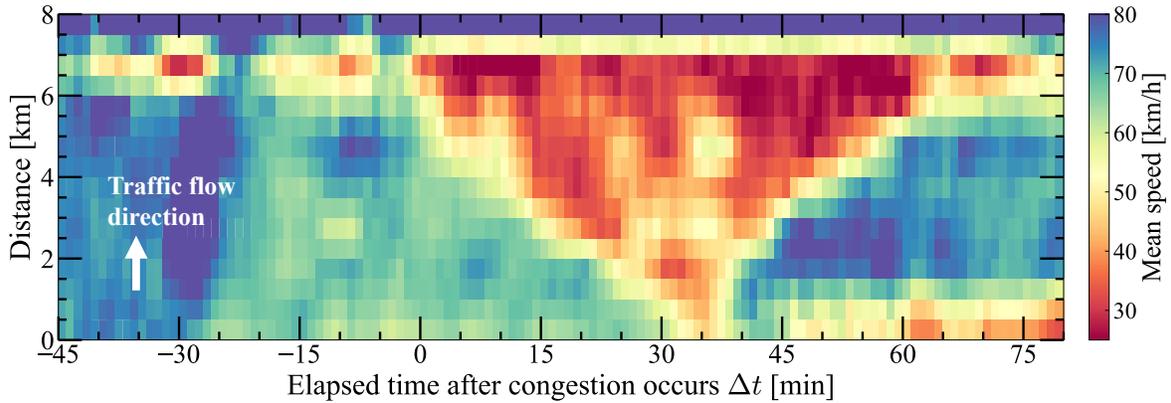

**Figure 5 Observed mean speeds in 500-meters, 1-minute patches obtained by DFOS on Route E85.**

In the theoretical model in the traffic simulation, this paper adopts the Stochastic Nishinari–Fukui–Schadschneider (S-NFS) model (*24*) as well as our previous work (*21–23*). It is a comprehensive stochastic cellular-automata-based model, which considers vehicle behaviors like drivers' recognition delay, anticipation, and involuntary deceleration. Thus, this model can reproduce realistic traffic-flow phenomena observed on actual freeways. The traffic simulator is self-developed and implemented in C++ and Python. Our data assimilation method estimates the optimal values of the probabilistic vehicle-behavior parameters in the S-NFS model so that the simulation result matches the observed mean speeds.

The lane-changing and overtaking behavior are also considered in the simulation. At each time step, each vehicle checks the achievable speed when it moves the adjacent lane considering the distances to surrounding other vehicles. If the achievable speed is higher when the vehicle moves to the adjacent lane, the lane change occurs at a certain probability. The lane change probability is set to be 10%, which is determined to reproduce a realistic lane-change behavior reported by the previous studies of field investigations (*25*, *26*).

The detailed simulation settings are followed by our previous work (*22*). The spatial resolution (cell length) is 10 meters, and the time step-size is 1.8 seconds. These settings are the same as the primitive study about the cellular-automata traffic flow model (*27*). Route E85 has two lanes with the nominal speed limit of 80 km/h. Therefore, the speed limit for the slow and passing lane without any controls in the simulation is 80 and 100 km/h, respectively.

This paper takes VSLs as an example of ATDM scenarios. Since VSL control effectiveness can be seen in the entire road, it is a suitable control scenario for the application of spatially continuous mean speeds by DFOS. To reproduce the VSL situations in the simulation, the speed limit in the range from the beginning point of the road to the bottleneck is changed. The four scenarios of VSLs are considered in total as shown on the left of **Figure 6**. The first one is no control. The second one is that the VSL is applied only to the passing lane. This scenario considers the control to reduce the gap in occupancy between the lanes. The third one is that the speed limit is reduced by 20 km/h. The last one is the combination of the reduced speed limits and the gap between the lanes. This paper refers to these four scenarios as "no control", "passing-lane (PL) VSL", "all-lanes (ALs) VSL", and "all-lanes (ALs) VSL2", respectively. Note that such lane-based VSL is not carried out on the validation freeway in practice. Evaluations of ATDM effectiveness in this paper is a "what-if" simulation to obtain implications before social implementation.

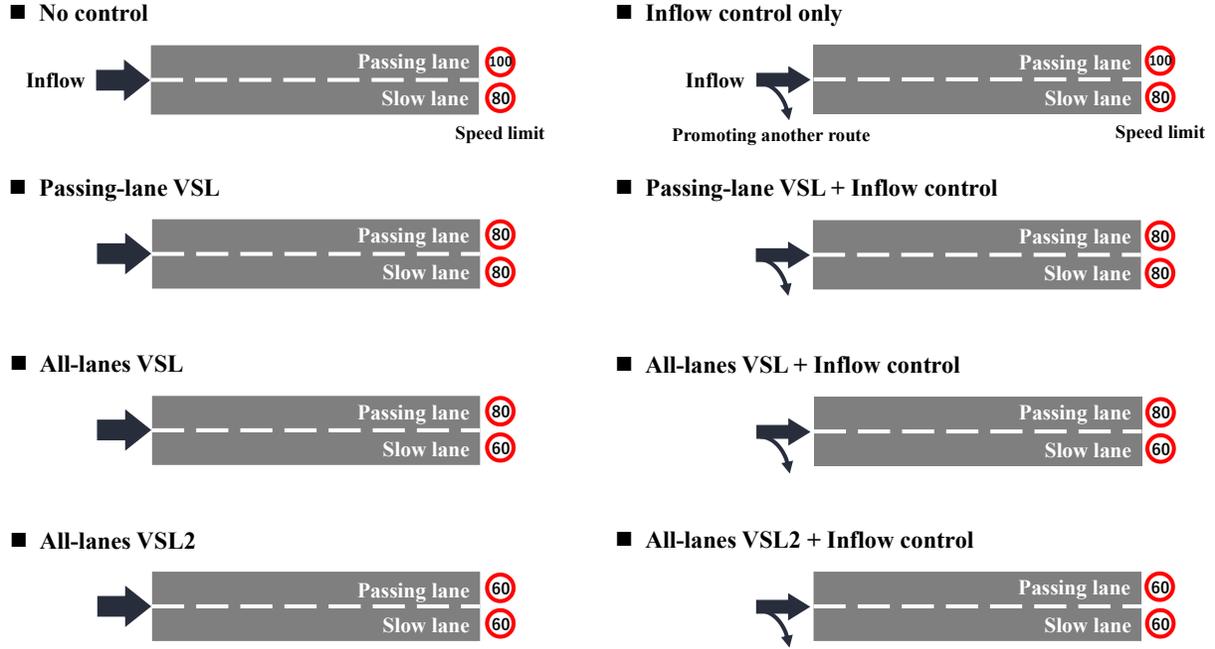

**Figure 6 Traffic control scenarios considered as examples of ATDM in simulation.**

In addition to these conventional VSLs above, this paper also considers the control for inflow as a result of TDM scenarios such as variable tolling. Specifically, inflow is reduced by promoting another route for drivers. For instance, in the case of Route E85 used for the validation, National Route 1, which exists parallel to Route E85, can be used as another route. With the combination of VSLs scenarios, the four scenarios are also considered as the traffic control as shown in the right of **Figure 6**.

Control scenarios of inflow are derived as follows. Firstly, the future inflow is predicted based on both short-term and long-term trends. The short-term prediction is based on the extrapolation of observed traffic volume in the past 30 minutes as shown in our previous work (*22*). The long-term one is based on the auto-regressive integrated moving-average (ARIMA) model. The ARIMA model is developed using the past 14-days inflow data obtained from the traffic counter. The resultant inflow without control is predicted as the average of the short- and long-term trend. The controlled inflow scenarios are created by considering that inflow is decreased in a quadratic manner with respect to time for simplicity. The quadratic model also considers inertia of controlled inflow because it is unrealistic to suddenly reduce inflow from the past. The controlled inflow scenarios are derived based on **Equation 3**,

$$Q_{\text{in,ctrl}}(t) = Q_{\text{in}}(t) + at + bt^2 \quad (a, b: \text{const.}) \qquad (3)$$

where $t$ is the elapsed time from the current time and $Q_{\text{in}}(t)$ is predicted inflow without any control.

The parameters $a$ and $b$ in **Equation 3** are determined as follows. The upper limit of $a, b$ is zero because this paper considers congestion mitigation by decreasing traffic demand. The lower limit is determined as below. At the late state of congestion since 30 minutes has passed since congestion occurred where traffic volume decreases to 35 veh/min, the controlled inflow in the next 30 minutes at the time does not become zero or less when $a \geq -0.3$ veh/min² and $b \geq -0.03$ veh/min³. Therefore, $(a, b) = (-0.3 \text{ veh/min}^2, -0.03 \text{ veh/min}^3)$ is set to be the lower limit. The increment of them is determined so that the increment is sufficiently large to ensure differences in results of each control scenario. Specifically, $a$ is sampled in the range from 0 to $-0.3$ veh/min² in the increment of $-0.025$ veh/min² (13 patterns) and $b$ ranges from 0 to $-0.03$ veh/min³ in the increment of $-0.005$ veh/min³ (7 patterns); 91 scenarios of the inflow control are considered. **Figure 7** shows several examples.

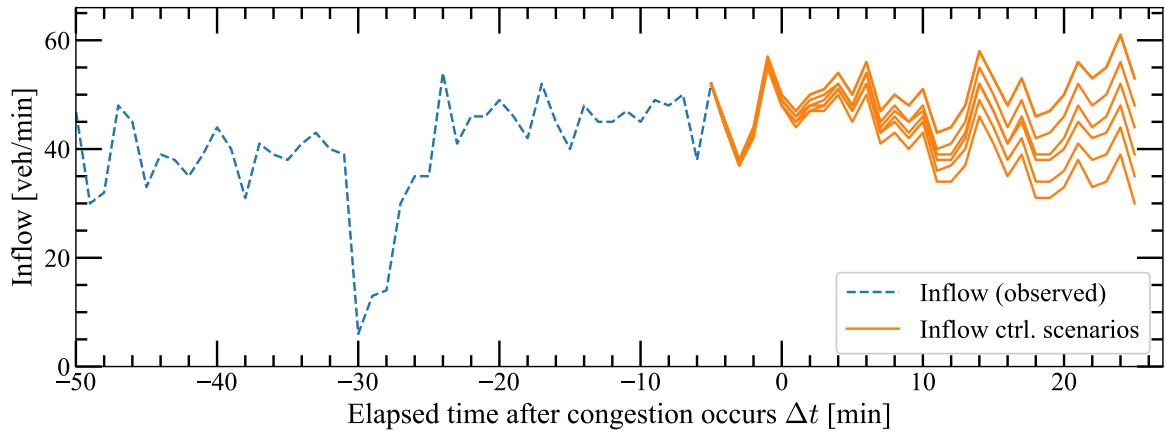

**Figure 7** Examples of inflow control scenarios. Control parameters are $a = 0, -0.15, -0.3$ veh/$\text{min}^2$ and $b = 0, -0.03$ veh/$\text{min}^3$ ($a$ and $b$ are defined in Equation 3).

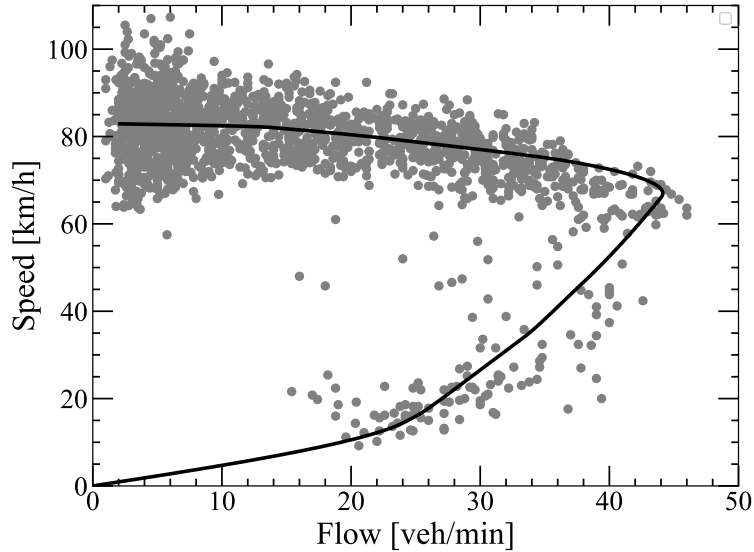

**Figure 8** Speed–flow diagram obtained on Route E85.

The optimal control scenario is estimated among these traffic control scenarios described above. To evaluate conventional ATDM approaches, this paper validates the traffic control effectiveness of VSLs without and with the inflow control scenarios. VSLs are widely implemented for ATDM, however, the inflow control is not always successful in promoting drivers' route choices. It also has a forcible aspect when the road operator precisely executes the inflow control scenario. Therefore, this paper regards traffic controls with only VSLs as a practical scenario and VSLs with the inflow control as an ideal scenario.

The objective functions are throughput and mean speed averaged over the validation road section. Throughput is a fundamental quantity to evaluate operational capability for transportation. Mean speed can be considered as a comprehensive quantity for the following reasons. Travel time and delay are also common traffic performance indices regarding congestion level. Nevertheless, they are approximately estimated as (road section length)/(mean speed); thus, they are related to mean speed. Congestion level such as occupancy is also correlated to mean speed via the density–speed diagram. Even for exhaust emission, it is a function of engine speed and travel time. Since they are related to driving speed, exhaust emission correlates with mean speed. For these reasons above, mean speed can is a representative of many traffic indices. Moreover, throughput and mean speed are competitive each other in congestion, which is a suited problem in the multi-objective optimization. Hence, this study adopts throughput and mean speed as objective functions.

Since the observed data from DFOS is only mean speeds, throughput is derived from the speed-flow diagram obtained from eight traffic counters on the validation section of the road as shown in **Figure 8**. The regression model denoted as the solid line is derived as follows. Firstly, the medians of speed within the bins of the 10 veh/km width along the density axis on the speed–density diagram. Secondly, the regression model for the speed–density diagram is derived using these medians. Finally, the regression model for the speed–density diagram is derived from that for the speed–density diagram.

Traffic flow where each control scenario is implemented is simulated up to 30 minutes from the present. Since effectiveness of control scenarios emerges approximately 10–20 minutes after implementation of the control scenario. Hence, it is sufficient to predict effectiveness of each control scenario up to 20–30 minutes ahead. From simulation results, throughput is derived from the mean flow rate using the Edie's method (*28*). The optimal controls of the two types of orientation, throughput-oriented and speed-oriented cases, are estimated. As mentioned in the previous section, they are estimated by adopting $(w, p) = (0.7, 1)$ and $(0.3, 1)$ in **Equation 1**, respectively. Optimal control scenarios are estimated every 5 minutes prior to 15 minutes before congestion occurs assuming that the control scenario is implemented for the first time at the time. The standard values to scale throughput and mean speed are 44.14 veh/min and 90 km/h, respectively. The former is determined by **Figure 8**.

**RESULTS & DISCUSSION**

**Only VSL Cases**
First of all, the robustness and reliability of our data assimilation results are validated. The accuracy of data assimilation and predictions of control effectiveness can be evaluated by comparing with the simulation result of the "no control" scenario and actual observed data. Accordingly, the mean percentage error (MPE) of predicted throughput and mean speed in 20–30 minutes are derived. Our data assimilation method achieves the accuracy with the MPE of 2.8% and 7.2% for throughput and mean speed, respectively. Therefore, the predicted effectiveness of each control scenario is reliable within the error margin.

Prior to estimating the optimal control, the effectiveness of each control scenario as a function of time is investigated. **Figure 9** shows predicted throughput and mean speed in 20–30 minutes when each control scenario is executed at each time. For example, **Figure 9 (a)** indicates that if no control is executed 15 minutes before congestion occurs, throughput is predicted to be 36.3 veh/min. At this moment, the PL VSL scenario is the optimal control to maximize throughput.

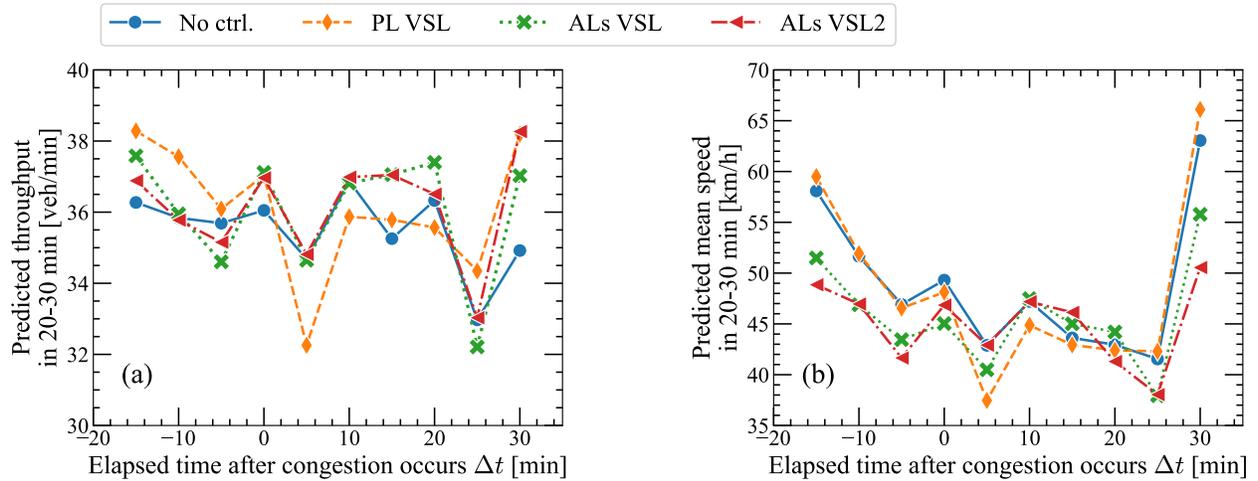

**Figure 9 (a) Predicted throughput and (b) mean speed as a function of elapsed time when each control scenario is executed at each time. Markers indicate each control scenario.**

The control scenario that maximizes throughput varies depending on time. Before congestion occurs, PL VSL is the most effective in maximizing throughput. In contrast, optimal control is ALs VSLs and ALs VSL2 when congestion occurs ($\Delta t \geq 0$ minutes). This tendency indicates that these VSLs suppress congestion and stabilize traffic flow as previous studies suggested (e.g., *2–5*). Optimal control to maximize mean speed shows a different tendency. Before congestion occurs, the no control scenario tends to be optimal control in addition to the PL VSL scenario. As time passes, ALs VSL and ALs VSL2 are the optimal control while congestion is growing as well as the maximizing throughput case. Prevention of growing congestion by the reduced speed limit for all lanes is beneficial for not only throughput but also mean speed. However, the achieved mean speed clearly decreases as time passes. The difference in the effectiveness between ALs VSL and ALs VSL2 is insignificant compared with that between the no control scenario and PL VSL. The reason for this result is that the frequency of lane-changing behavior in the reduced speed limit condition is lower than that in the other condition.

*Y. Yajima, H. Prasad, D. Ikefuji, T. Suzuki, S. Tominaga, H. Sakurai, and M. Otani*

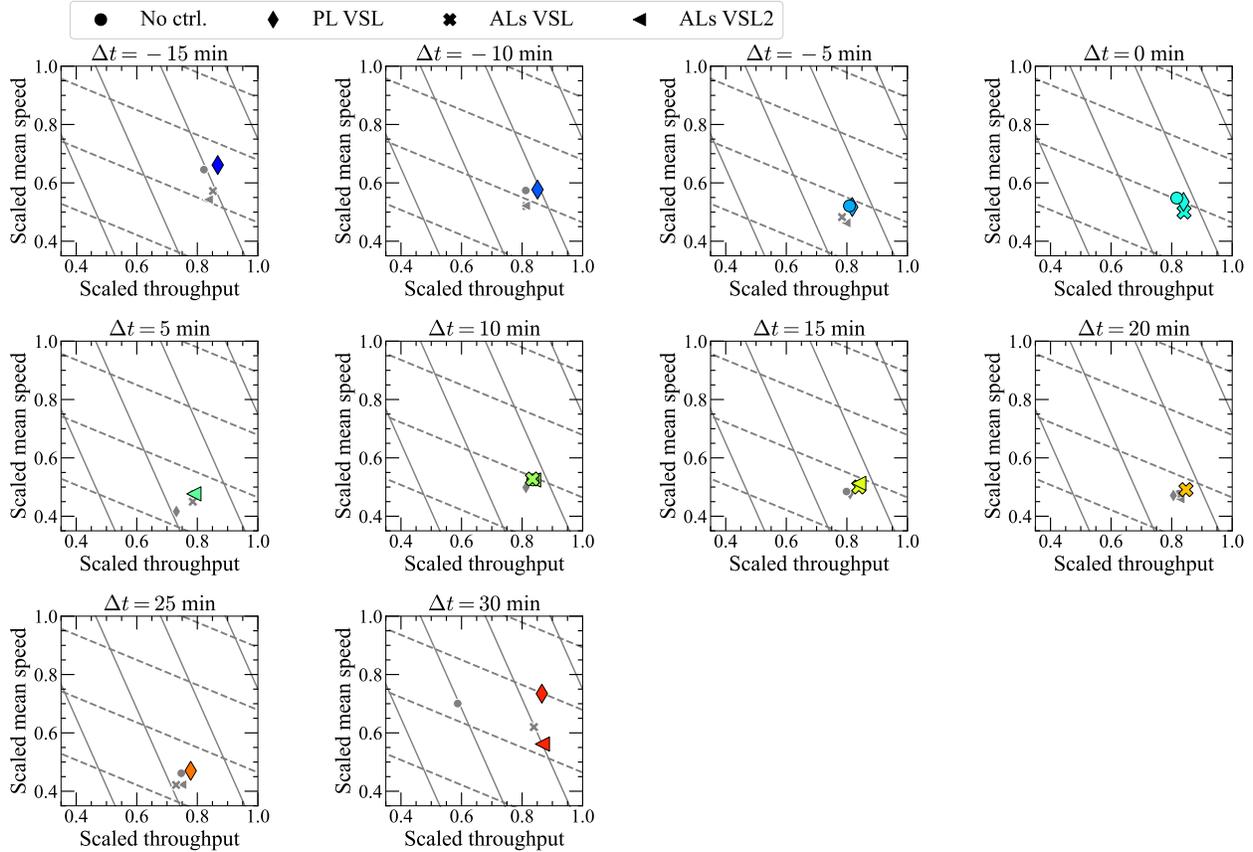

**Figure 10** Distributions of Pareto solutions on the objective-functions space at each time. Pareto and non-Pareto solutions are denoted as colored and gray markers, respectively. Colors and shapes of markers indicate time and each control scenario, respectively. Diagonal solid and dashed lines are contours of norm at $(w, p)$ = (0.7, 1) and (0.3, 1) corresponding throughput- and speed-oriented scenario, respectively.

**Figure 10** shows the distributions of the Pareto and non-Pareto solutions at each time. When the inflow control is not considered, the Pareto solutions are few; therefore, optimal control is limited options. Among the Pareto solutions, the throughput- and speed-oriented optimal solutions are identified using **Equation 1**. **Figure 11** shows the predicted throughput and mean speed in 20–30 minutes at optimal control for the throughput- and speed-oriented scenarios. Their improvement rate by optimal control compared with the no control scenario is also shown. For example, **Figure 11** indicates that the PL VSL scenario is optimal control at $\Delta t = -15$ minutes for throughput- and speed-oriented optimization. Since the Pareto solutions are a few, the resultant optimal solutions for the throughput- and speed-oriented scenarios coincide with each other.

In both orientation scenarios, the PL VSL scenario is optimal control when congestion does not occur yet ($\Delta t \leq -10$ minutes) and it will turn into dissipation in several minutes ($\Delta t \geq 25$ minutes) as shown in **Figure 9** (cf. congestion turns into dissipation at $\Delta t \sim 35$ minutes, see **Figure 5**). Its main reason is that the usage rate (i.e., occupancy) on each lane is balanced owing to the same speed limit. Congestion tends to occur due to the unbalanced occupancy caused by the drivers' attempt to move to the passing lane in order to drive faster (*29–34*). In fact, simulation results show that the usage rate on the passing lane decreases from ~65% to ~50% when compared with the PL VSL and no control scenario. The explicit VSL only on the passing lane restrains drivers from moving to and staying in the passing lane. As a result, traffic flow remains smooth and a higher throughput than that in the no control scenario is retained. This result implies that the lane usage is important as suggested by the previous study (*32*).

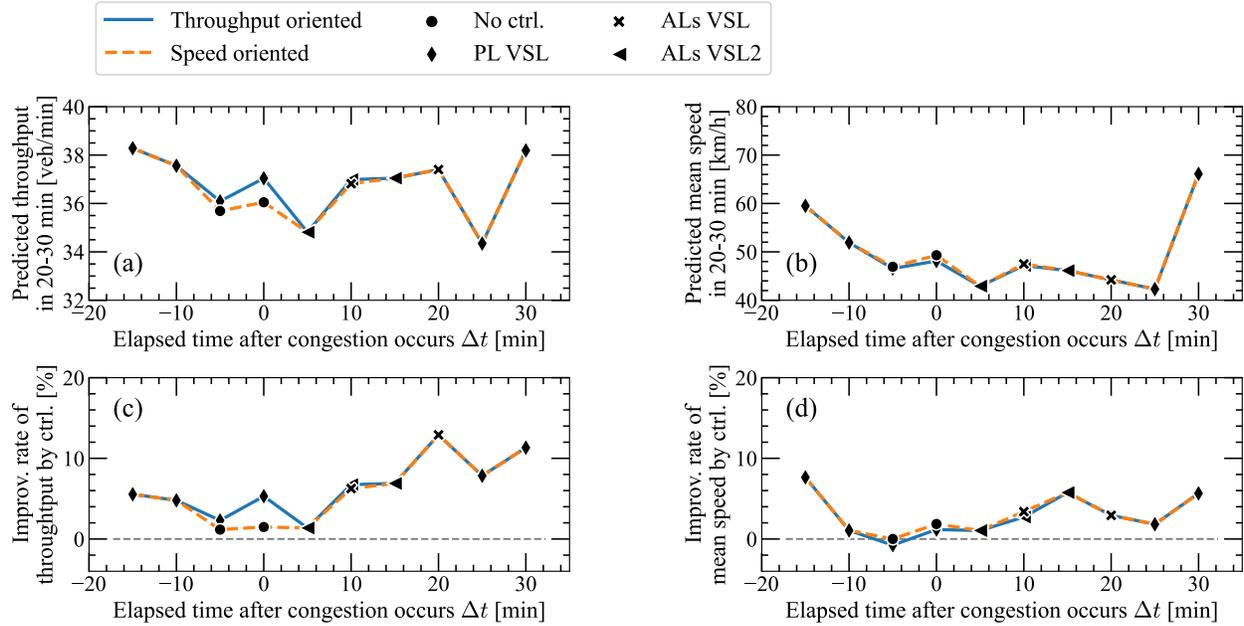

**Figure 11 (a) Predicted throughput in 20–30 minutes in the throughput- and speed-oriented optimal control scenario as a function of time. (b) The same as (a) but for mean speed. The panels (c) and (d) show the improvement rate of throughput and mean speed by the optimal control compared with the no control scenario. Shapes of markers indicate which scenario is optimal control.**

In the case that growing congestion accounts for a large part within the prediction horizon ($0 \leq \Delta t \leq 20$ minutes), ALs VSL and ALs VSL2 become the optimal control scenario. This is because the number of vehicles reaching the tail end of congestion decreases as previous studies suggested (*2, 4*). As a result, congestion does not get worse compared with the no control scenario.

Moreover, **Figure 11 (c)** indicates that throughput can be recovered regardless of the control execution time whereas predicted throughput and its improvement rate fluctuate due to the continuously varying traffic state. Even if the optimal control is executed after 10–20 minutes have passed since congestion occurs, throughput can be recovered and improved compared with the no control scenario. The improvement rate of throughput is ~5% shortly before and after congestion occurs ($\Delta t \leq 5$ minutes); furthermore, it is still as high as ~8%, at most 14% in $\Delta t > 5$ minutes.

A possible reason for recoverable throughput even after congestion occurs is as follows. Since the capacity drop phenomenon (*35*) lasts until congestion completely dissipates, throughput continues to decrease without any control. Therefore, a slight improvement or just keeping throughput results in an improvement rate of several percent. In addition, the speed–flow diagram shown in **Figure 8** provides an interpretation. According to the diagram, the recovered throughput of 35–40 veh/min by optimal control corresponds to the mean speed of 40–50 km/h. On the other hand, in the case of the no control scenario, predicted throughput and mean speed are ~30 veh/min and 20–30 km/h, respectively. Hence, the recovery of throughput would be practically promising because all we need through VSL is to execute optimal control that increases mean speed from ~30 km/h to ~40 km/h in the next 20–30 minutes. To prevent mean speed from decreasing to ~30 km/h, the VSL control alone is adequate. Since the speed–flow diagram is universal for traffic flow, this trend might be applicable on other freeways while specific numbers of throughput and mean speed depend on roads and congestion states.

In contrast to throughput, the mean speed achieved by optimal control gradually decreases as time passe. The improvement rate of mean speed also decreases. Since inflow is not controlled, traffic demand keeps increasing in the beginning stage of congestion. In addition, the VSL scenarios cannot decrease occupancy. Consequently, mean speed cannot be recovered even when optimal control is executed. At $\Delta t = 30$ minutes, since congestion is almost cleared by the prediction horizon, the predicted mean speed by optimal control significantly increases. These results indicate that if mean speed, i.e., express transportation, is prioritized, any traffic control scenario disregarding inflow control does not make the future traffic state approach the optimal state, unless congestion is almost cleared by the prediction horizon. TDM is important to maintain mean speed as high as possible. If the mean speed is required to remain, the prior control of the PL VSL scenario should be executed at least 15 minutes before congestion occurs.

In the ALs VSL and ALs VSL2 scenarios, dispersion of mean speed at each road segment decreases. The simulation results show that the standard deviation of predicted mean speed in the AL VSL and AL VSL2 scenarios is lower by 58% than that in the no control scenario. This suppressed speed dispersion is expected to decrease the risk of accidents as reported by the previous studies of VSL (e.g., *1–3, 36*) although express transportation cannot be achieved as mentioned above.

Finally, the assessment of input data to the reliability of the optimal control estimation is examined likewise in the first paragraph of this subsection. The prediction accuracy when data assimilation is carried out without DFOS is also investigated. Our previous work (*21*) reveals that when traffic counters are used instead of DFOS for data assimilation, the optimal model-parameter estimation is failed due to the intermittent traffic data. As a result, the prediction error approximately doubles compared with when using DFOS. In the case of this study, the MPE of the predicted throughput and mean speed is expected to be 5.6% and 14%, respectively, according to the first paragraph of this subsection. These errors are comparable to the predicted improvement rate. Thus, the performance of the optimal control estimation without DFOS is no longer reliable.

**VSL & Inflow Control Cases**
**Figure 12** shows the distributions of Pareto solutions. In contrast to **Figure 10**, many Pareto solutions exist when considering inflow control, which indicates that there are more optimal options. Since both objective functions are maximized, the distribution of Pareto solutions is convex to the top right on the objective functions space. Distributions of the Pareto solutions for each traffic control scenario show a rough trend. Only inflow control scenarios show increased mean speed and decreased throughput. The solutions of the PL VSL + inflow control scenarios are distributed in the regime shifted to the right along the axis of throughput from the only inflow control scenarios. The solutions of the ALs VSL2 + inflow control scenarios have higher throughput and lower mean speed than those of the PL VSL + inflow control scenarios. The solutions of the ALs VSLs + inflow control scenarios are distributed in a lower throughput and lower mean speed compared with other control scenarios. However, these tendencies become ambiguous as congestion gets serious, especially at $\Delta t = 20$ minutes. At the time, some of the ALs VSL + inflow control scenarios become the Pareto solutions.

The optimal controls of the throughput- and speed-oriented scenarios are estimated in the same way as described in the previous subsection. **Figure 13** shows which scenario is optimal control and its improvement rate at each time compared with the no control scenario. The balance-oriented optimal controls derived from $(w, p) = (0.5, 2)$ as shown in **Figure 4 (c)** are between throughput- and speed-oriented optimal solutions. The significant difference from **Figure 11** is that higher throughput and mean speed are achieved by optimal control in each orientation scenario. In the throughput-oriented case, the optimal control improves throughput by ~10–15%. In the speed-oriented case, mean speed is significantly improved by 20–30%, 38% at most. Moreover, the improvement rate of throughput in the speed-oriented case and that of mean speed in the throughput-oriented case is higher than zero. The reason for these enhanced consequences is that inflow control can suppress the increased occupancy. In fact, the simulation results show that the density decreases by 19% on average compared with the optimal control scenario disregarding the inflow control scenarios.

These improvement rates are comparable to ATDM frameworks reported by previous studies. For example, numerical experiments indicate that VSL improves the total travel time by ~10–20% (*10*, *37–39*). Regarding another ATDM scenario, the dynamic part-time shoulder use recovers mean speed by ~50% (*40*), which is much higher improvement rate than the result in **Figure 13 (d)**. This significant effectiveness may indicate that ATDM scenarios directly changing traffic capacity result in effective control. Note that the improvement rates between this study and these previous works are not comparable because premises such as used data, traffic state, and the road model for validation are different.

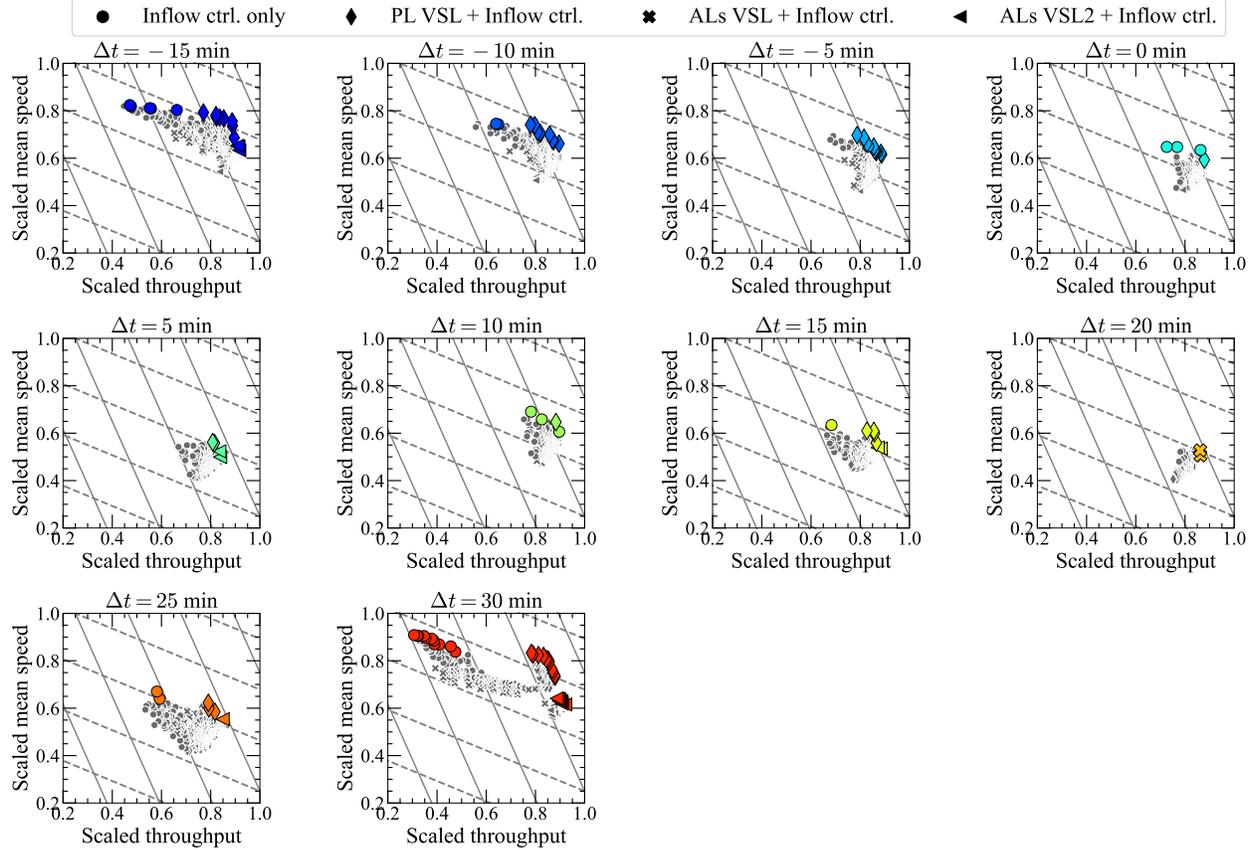

**Figure 12 The same as Figure 10 but for the case considering inflow control scenarios.**

The PL VSL + inflow control scenarios are often optimal control as is the case of the VSL alone, which indicates that balancing the usage rate among lanes is also effective. In addition, ALs VSL and ALs VSL2 with inflow control are the optimal control scenarios only when congestion is getting serious by the prediction horizon ($\Delta t = 5, 20,$ and $25$ minutes) as seen in **Figure 13** while this tendency can be seen for the throughput-oriented scenario. In the speed-oriented scenario, the ALs VSL and ALs VSL2 scenarios are not chosen as optimal control because mean speed is lower than that in other scenarios due to the reduced speed limit for all lanes.

Moreover, there is an advantage to proactive control before congestion occurs. The parameters $a$ and $b$ in **Equation 3** for the optimal controls gradually decrease as time passes. Specifically, $(a, b) = (-0.075 \text{ veh/min}^2, -0.020 \text{ veh/min}^3)$ for both throughput- and speed-oriented optimal control at $\Delta t = -15$ minutes, however, $(a, b) \simeq (-0.275 \text{ veh/min}^2, -0.030 \text{ veh/min}^3)$ in both throughput- and speed-oriented scenarios when congestion shortly occurs or already occurs. Namely, the future inflow in 20–30 minutes needs to be reduced by 26 veh/min in $\Delta t > 0$ minutes although inflow control at $\Delta t = -15$ minutes only requires that the inflow has to be reduced by 15 veh/min. The amount of traffic volume to promote another route has to be increased when congestion occurs in the next 10 minutes or already occurs; the demand management becomes stern.

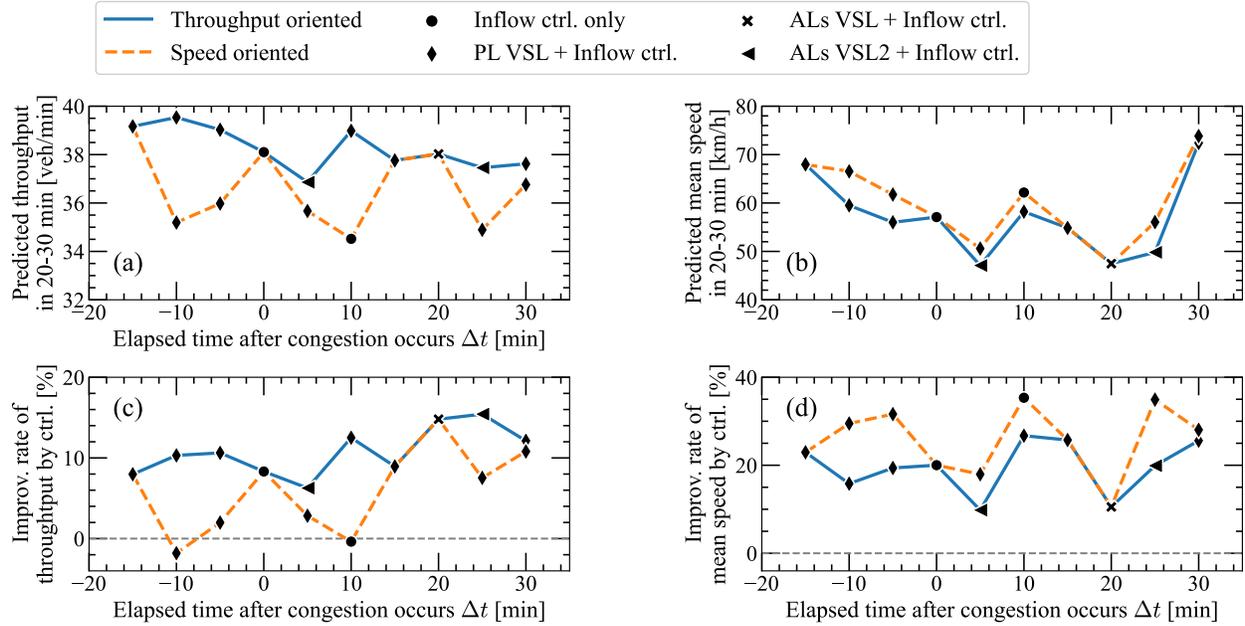

**Figure 13** The same as Figure 11 but for the case considering inflow control scenarios. Note that the vertical scale of the panel (d) is different from that in Figure 11.

## CONCLUSIONS

This paper has presented the real-time optimal-control estimation for the current traffic state using DFOS and the data assimilation method. In terms of the multi-objective optimization, this paper has also provided the estimation method to uniquely determine the optimal solution from the Pareto solutions. Taking the VSL and inflow control as examples of traffic control, our method is validated using congestion data obtained by DFOS on a Japanese freeway. Optimal controls for maximizing throughput and mean speed are estimated. Our data assimilation method successfully predicts throughput and mean speed in 30 minutes with the MPE of 3–7%. The main findings and implications are described as follows.

i). When VSLs alone are considered and inflow control is disregarded (i.e., a practical case), the explicit VSL for the passing lane tends to be optimal control to maintain throughput as high as possible when congestion does not occur yet, it is in the early stage, and it is almost cleared. This VSL balances the occupancy among lanes; as a result, transportation efficiency increases. When congestion is serious, the VSL reducing the speed limit for all lanes is optimal control. However, it is impossible to keep mean speed; it decreases as time passes even when the speed-oriented optimal control is executed. To maximize both throughput and mean speed by only the VSL, proactive control is necessary before congestion occurs, at least prior to 15 minutes. The improvement rate of throughput by optimal control achieves 5–14% whereas that of mean speed is less than 8% without the proactive control.

ii). When both VSLs and inflow control are considered (i.e., an ideal case), the optimal control can improve throughput and mean speed by 10–15% and 20–30%, respectively, compared with only the VSLs case. This high improvement rate is attributed to the decreased occupancy by inflow control. As well as the VSLs alone case, the VSL reducing the speed limit for all lanes is effective in maximizing throughput when congestion is serious. Otherwise, VSL for only the passing lane is often throughput-oriented optimal control. In addition, the inflow restriction can be mild when proactive control is executed.

In summary, this study indicates the importance of proactive control before congestion occurs in any optimization scenarios. For the advanced ATDM, the proactive optimal control estimation will be a keen subject.

**AUTHOR CONTRIBUTIONS**
The authors confirm contribution to the paper as follows: study conception and design: Y. Yajima, H. Prasad, D. Ikefuji, T. Suzuki, S. Tominaga, H. Sakurai; data collection: D. Ikefuji, T. Suzuki, S. Tominaga, M. Otani; analysis and interpretation of results: Y. Yajima, H. Prasad, D. Ikefuji; draft manuscript preparation: Y. Yajima, H. Prasad, D. Ikefuji, T. Suzuki, S. Tominaga, H. Sakurai, M. Otani. All authors reviewed the results and approved the final version of the manuscript.

**Declaration of Conflicting Interests**
The authors declared no potential conflicts of interest with respect to the research, authorship, and/or publication of this article.

**Funding**
The authors disclosed no financial support for the research, authorship, and/or publication of this article.